\documentclass[a4paper]{VisionStyle}
\usepackage{epsfig}

\begin{document}

\title{Spectral response of the EPIC-pn detector: basic dependences}

\author{F.\,Haberl \and U.G.\,Briel \and K.\,Dennerl \and V.E.\,Zavlin } 

\institute{  Max-Planck-Institut f\"ur extraterrestrische Physik,
             Giessenbachstra{\ss}e 1, 85748 Garching, Germany          }

\maketitle 

\begin{abstract}

One of the three cameras behind the X-ray telescopes on board XMM-Newton is 
equipped with a pn CCD detector. Here the current status of the pn spectral 
response calibration is summarized. 
Several parameters describing the spectral response of the EPIC-pn 
detector show spatial dependencies. E.g. spectral resolution
and single- to double-pixel event ratios are functions of the
charge transfer losses and therefore depend on the distance to
the read-out node of the CCD (the RAWY detector coordinate).

\keywords{Missions: XMM-Newton -- Instruments: EPIC-pn -- 
                                  Instruments: calibration}
\end{abstract}

\section{Introduction}
  
On December 10, 1999 the XMM-Newton X-ray observatory 
(\cite{fhaberl-WA2poster:ja01}) was launched into a 48 h 
Earth orbit by an ARIANE V rocket. One of the three European Photon Imaging
Cameras (EPIC, \cite{fhaberl-WA2poster:tu01}) is equipped with a CCD detector
based on pn technology (\cite{fhaberl-WA2poster:st01}). Ground measurements 
and modeling of the spectral detector response of the EPIC-pn camera are
described in \cite*{fhaberl-WA2poster:po99} and first results on the in-orbit
performance of the EPIC-pn camera are presented in 
\cite*{fhaberl-WA2poster:br00}. The current status of the EPIC-pn calibration 
is reported by \cite*{fhaberl-WA2poster:br02} in these conference proceedings.

\section{Spectral response of the pn-CCD detector}

The spectral detector response matrix (DRM) describes the detection probability 
for an X-ray photon with given energy in each of the 4096 ADU channels of the pn 
detector. The response to X-rays is calculated using the Partial Event
Model (PEM) which is based on the charge collection function 
(\cite{fhaberl-WA2poster:po99}). 

\begin{figure}[ht]
  \begin{center}
    \epsfig{file=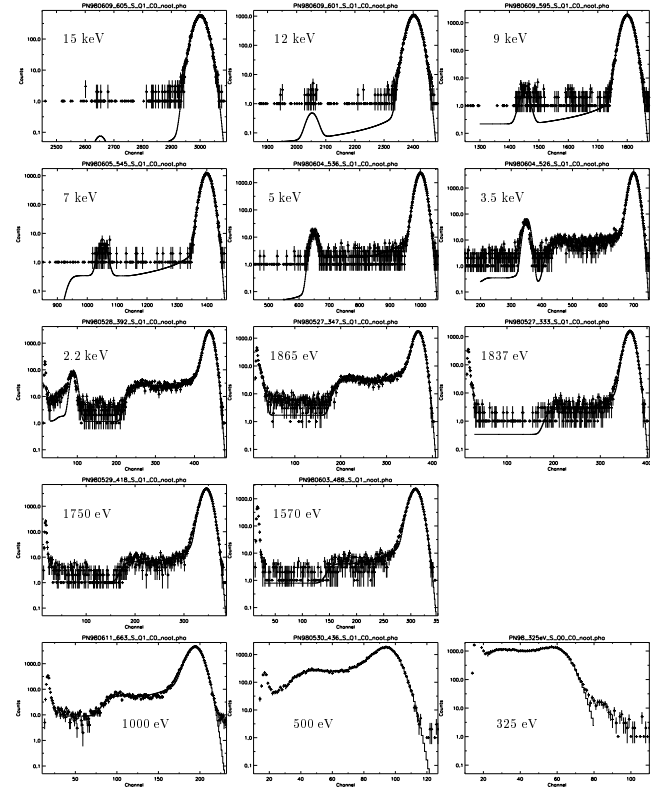, clip=, angle=0, width=85mm}
  \end{center}
\caption[]{Spectral response to monochromatic radiation for 14 different
energies as measured at the IAS synchrotron facility in Orsay, France, using the
spare detector module (FM1). The spectra are averaged over CCD0 of quadrant 1
and shown with logarithmic intensity scale.
The shoulder and flat shelf at the low energy side of the main peak are visible.
Note the jump in shoulder height at the Si edge. At higher energies the Si escape peak is also
visible.}  
\label{fhaberl-WA2poster_fig:fig1}
\end{figure}

\begin{figure*}[ht]
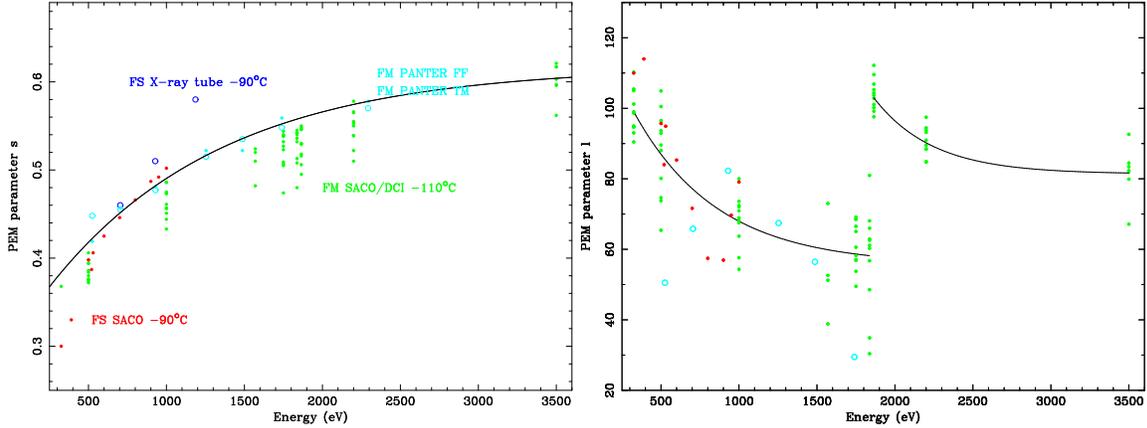

  \begin{center}
    \epsfig{file=fhaberl-WA2poster_fig2a.ps, clip=, angle=-90, width=75mm}
    \epsfig{file=fhaberl-WA2poster_fig2b.ps, clip=, angle=-90, width=75mm}
  \end{center}
\caption[]{PEM parameters s and l which describe the shape
of the low-energy shoulder as function of energy. The lines define best fit
functions used in the DRM generation for the interpolation between energies.}  
\label{fhaberl-WA2poster_fig:fig2}
\end{figure*}

\begin{figure*}[ht]
  \begin{center}
    \epsfig{file=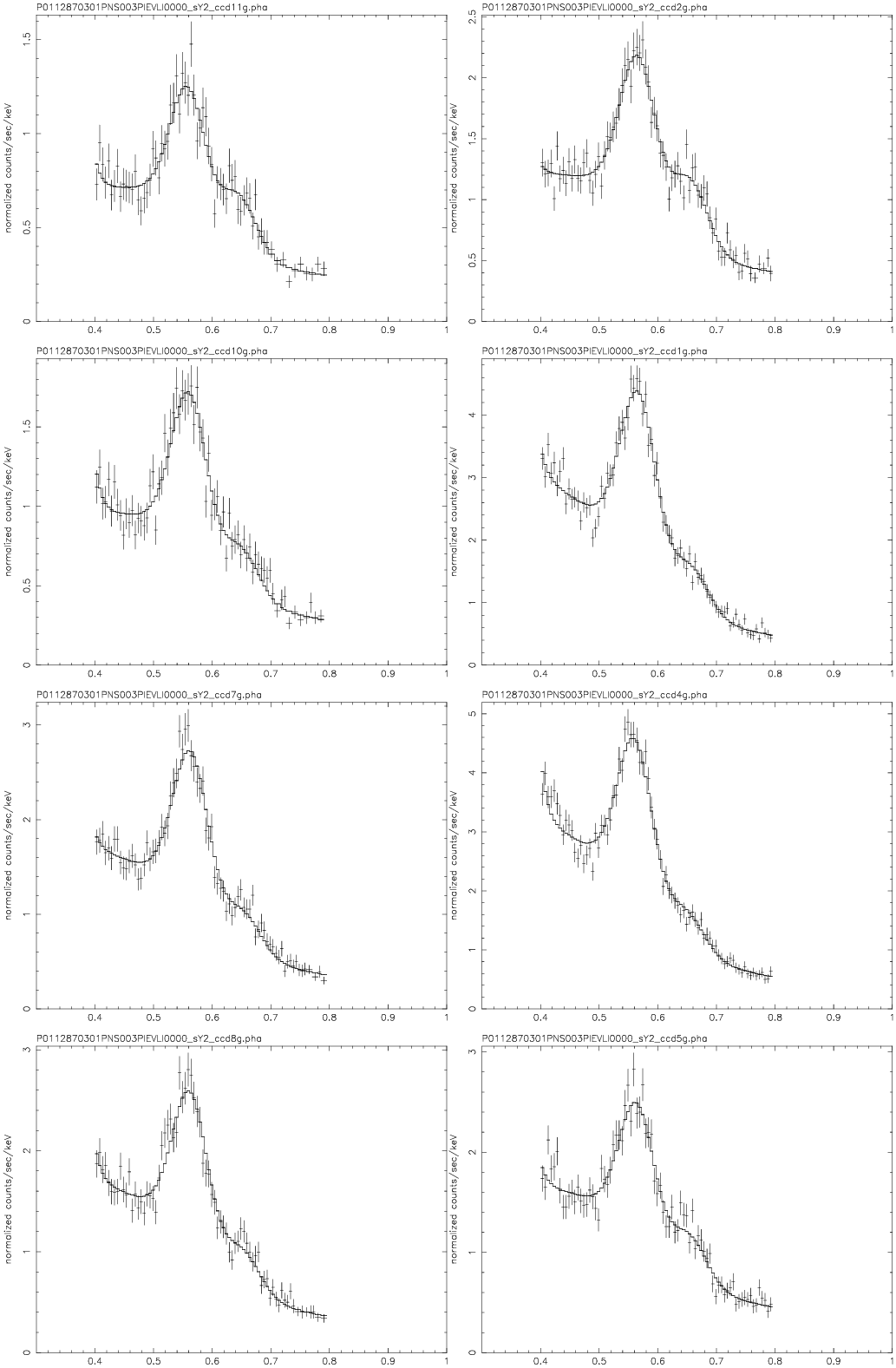, clip=, width=88mm}
    \epsfig{file=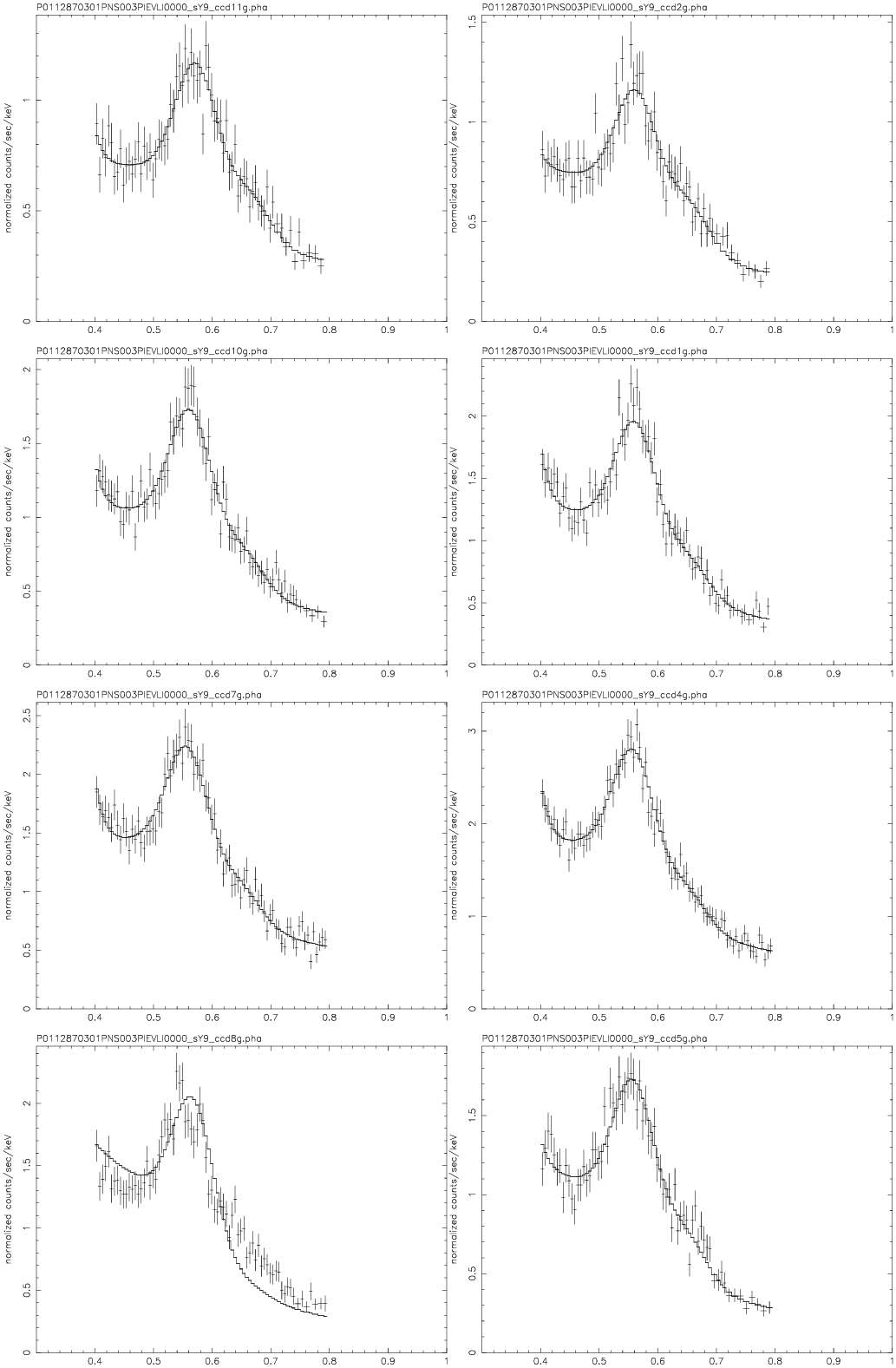, clip=, width=88mm}
  \end{center}
\caption[]{EPIC-pn spectra as function of channel energy (keV) 
obtained from extended fullframe mode observations of
the Vela SNR. The large extent of the SNR and the soft spectrum with the
dominating OVII He$\alpha$ line triplet allows to investigate spatial
dependences of the spectral response at low energies. Spectra were selected 
from CCD lines (RAWY) 20-79 (left) and from lines 180-199 (right) for the inner
four CCDs using single-pixel events. The OVII line blend is broader at 
larger distances to the read-out node.}  
\label{fhaberl-WA2poster_fig:fig3}
\end{figure*}

\subsection{Energy dependence}

The PEM accounts for the effects of incomplete charge collection when photons 
are absorbed close to the detector surface. Different processes are responsible
for the different spectral features that are visible in the measured spectrum.
Figure~\ref{fhaberl-WA2poster_fig:fig1} shows the spectral response to
monochromatic X-rays for various photon energies. Absorption in the surface
oxide causes detection of very few electrons that make it over the Si - SiO$_2$ 
boundary, resulting in a flat shelf that extends down to the noise peak 
(e.g. clearly visible in the 1 keV spectrum). If the photon is absorbed in the 
Silicon, but close to the
Si surface, only a part of the generated electrons will reach the read-out node
and a shoulder in the spectrum is produced. If the absorption takes place deep
in the detector, all generated electrons will be detected. Since the absorption
depth generally increases with energy the surface effects play a minor role at 
high energies (see e.g. the spectra from energies above 7 keV) and the 
response resembles a Gaussian peak with a width defined by
the total detector noise (Fano, charge transfer and electronic noise).

The ground calibration spectra were used to derive the parameters for the PEM.
Figure~\ref{fhaberl-WA2poster_fig:fig2} shows the energy dependence  
of the s and l parameters. The PEM parameter s describes the minimum fraction 
of electrons which are
detected when the photon is absorbed in the Si and is a direct measure for the
low energy ``end" of the shoulder. The parameter l is a characteristic length
scale for the depth of the transition region near the surface of the detector
where the charge collection function changes from minimum s to full charge 
collection of 1 which is typically reached after 50-100 nm. The parameter l
determines the height of the shoulder.

\subsection{Energy resolution}

The energy resolution of the detector is given by statistical processes in the
charge production due to photo absorption, in the charge transfer during read-out 
and in the detector electronics. The charge transfer inefficiency (CTI), caused
by traps in the Silicon which capture electrons, its modeling and 
correction is explained in detail in \cite*{fhaberl-WA2poster:de02} in these 
conference proceedings. Although the CTI losses can be corrected for, they 
introduce a noise component which influences the energy resolution. This noise 
increases with distance of the detector pixel in which the photon is recorded 
to the read-out node (the RAWY detector coordinate). 
Figure~\ref{fhaberl-WA2poster_fig:fig3} illustrates the degrading energy
resolution with increasing RAWY distance on spectra obtained in orbit. 

\begin{figure}[ht]
  \begin{center}
    \epsfig{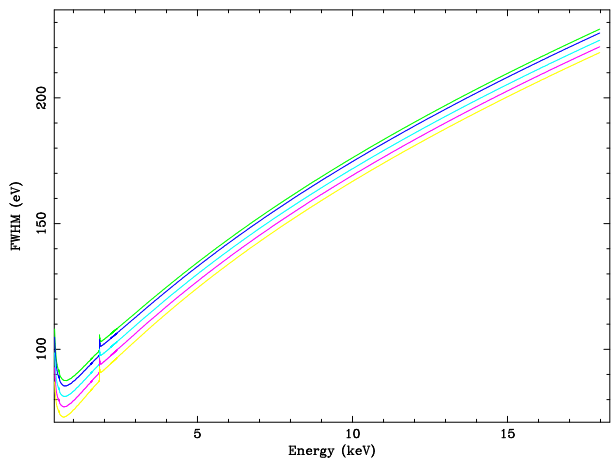}
  \end{center}
\caption[]{Spectral resolution of the EPIC-pn (single-pixel events) detector 
vs. energy for five line (RAWY) regions with different distance to the 
read-out node. Y1 (bottom curve, yellow) denotes RAWY=20-39, Y3 RAWY=60-79, 
Y5 RAWY=100-119, Y7 RAWY=140-159 and Y9 (upper curve, green) RAWY=180-199. 
The jump of the shoulder height (PEM parameter l) at the Si edge is 
reflected in the resolution. At the lowest energies the
shoulder dominates the spectrum.}
\label{fhaberl-WA2poster_fig:fig4}
\end{figure}

The energy resolution in terms of FWHM was derived
directly from the current version of the response matrix by determining the
width of the main spectral peak (for given photon energy) at half of the maximum
intensity. Figure~\ref{fhaberl-WA2poster_fig:fig4} shows the FWHM in eV as
function of energy for five different line regions in the detector.

\subsection{Single- and double-pixel events}

\begin{figure*}[ht]
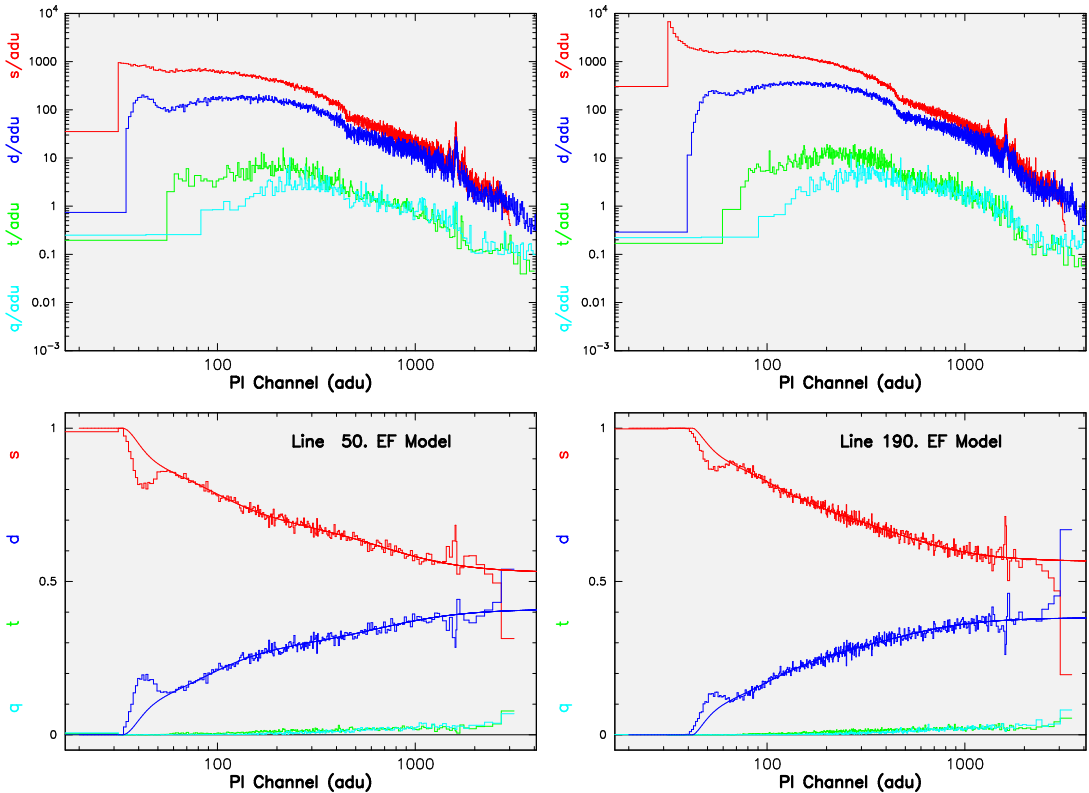

  \begin{center}
    \hbox{\epsfig{file=fhaberl-WA2poster_fig5a.ps, clip=, angle=0, width=72mm}
          \epsfig{file=fhaberl-WA2poster_fig5b.ps, clip=, angle=0, width=72mm}}
  \end{center}
\caption[]{EPIC-pn spectra of single-, double-, triple-, and quadruple-events
plotted as histograms (upper panels). The spectra were obtained from combined 
Coma observations in extended fullframe mode. In the lower panels the relative 
fractions of the different event types are shown vs. CTI-corrected PI channels. 
On the left spectra were selected from RAWY=40-59, on the right from
RAWY=180-199. Note the increasing effective event threshold with increasing 
RAWY. The SAS task epatplot is available to produce these plots.}  
\label{fhaberl-WA2poster_fig:fig5}
\end{figure*}

In the majority of cases an X-ray photon produces an electron cloud which is
confined to a single detector pixel. However in cases where the photon is absorbed
close to a pixel border the charge is split between two or more pixels (at 
most four in the case of the pn detector with its square pixel size of 150
$\mu$m),
producing double-, triple- and quadruple-pixel events. Because of the increasing
size of the electron cloud at higher X-ray photon energies, the fraction of 
higher order events increases with energy. In the upper panels of
Figure~\ref{fhaberl-WA2poster_fig:fig5} two examples of pn spectra are shown.
For each of the four valid event patterns a spectrum was binned. Double-event
spectra start at higher PI channels (corrected for CTI losses) because to be
recorded as ``double" both events must exceed the electronic low-energy event 
threshold
at the read-out node (which is currently set to 20 adu for the imaging modes 
and nominally 40 for timing mode). In a similar way this holds for ``triples" 
and ``quadruples". In the lower panels of Figure~\ref{fhaberl-WA2poster_fig:fig5} 
the relative fractions of the four valid event types normalized to their sum 
are plotted as function of energy. For the reason described above the 
single-fraction is 1.0 up to twice the threshold where doubles become 
detectable. However a small fraction of
these singles are so-called pseudo singles, in reality doubles for which one of
the two split partners fell below the event threshold. A further complication
is the RAWY dependence of the event threshold. The fixed value defined in the
electronics translates into effective thresholds which increase with distance
to the read-out node due to CTI losses. An event which produces charge just 
above the threshold can fall below after the charge transport. This is
demonstrated in Figure~\ref{fhaberl-WA2poster_fig:fig5} for spectra taken closer
to the read-out node (left) and near the detector center (right). Clearly seen
is the start of the double spectrum at higher PI values in the latter case.

Comparison of single and double fractions derived from in-orbit data have shown that
they are independent of source spectrum and source intensity as long as the spectrum
is free of pile-up. This allows to derive standard curves for single- and double-
fractions (triples and quadruples are neglected for spectral analysis due to their
much reduced spectral resolution and small fractions) for each read-out mode.
At higher intensities so called pattern pile-up increases the
double fraction due to the production of false double-pixel events which actually 
consist of two neighboring single-pixel events. These false doubles cause
deviations in the single and double fractions at PI channels typically twice the
channels where the spectrum has its intensity maximum (there the probability for
pile-up is highest). Examples are presented by \cite*{fhaberl-WA2poster:br02} 
and \cite*{fhaberl-WA2poster:ha02} in these proceedings.

A different kind of deviation from the standard curves at low PI channels 
can occur when the detector noise during an observation is at unusually high
levels or when a spectrum is extracted from a region close to the read-out node
where this noise is enhanced. 
A high level of low energy noise can result in an increased number of doubles,
i.e. a noise peak in doubles is created at twice the energy of the noise seen 
in singles. This noise peak strongly increases the fraction of doubles just
above twice the threshold (and correspondingly decreases the relative number of
singles) as seen in Figure~\ref{fhaberl-WA2poster_fig:fig5} by the deviations
of the fractions at low energies from the expected curves. For a spectral 
analysis only energies above this double noise peak should be used and it is
therefore recommended to produce pattern statistic plots from the events
in the source extraction region. 

\begin{figure}[ht]
  \begin{center}
    \epsfig{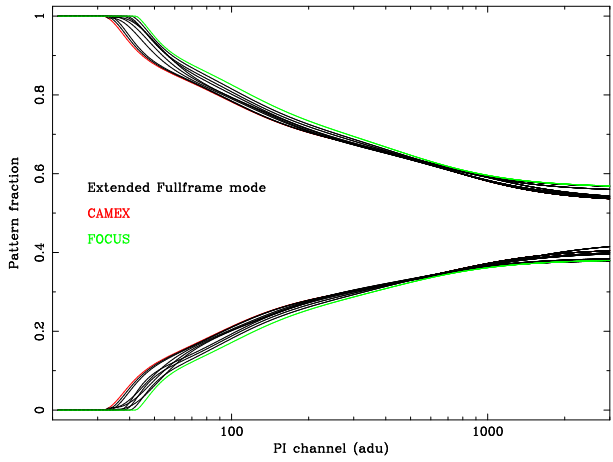}
  \end{center}
\caption[]{Standard curves for single (upper lines) and double fractions.
Each line was obtained from spectra with different RAWY detector coordinates.
Shown are ten curves from Y0: RAWY=0-19 (CAMEX) to Y9: RAWY=180-199.}  
\label{fhaberl-WA2poster_fig:fig6}
\end{figure}

The dependence of the effective event threshold and therefore the single and double
fractions on RAWY is demonstrated in more detail in 
Figure~\ref{fhaberl-WA2poster_fig:fig6} by plotting the standard curves derived from
areas with different distance to the read-out node (CAMEX). 

\begin{acknowledgements}

The XMM-Newton project is an ESA Science Mission with instruments and 
contributions directly funded by ESA Member States and the USA (NASA). The
XMM-Newton project is supported by the Bundesministerium f\"ur Bildung und
For\-schung / Deutsches Zentrum f\"ur Luft- und Raumfahrt (BMBF / DLR), the
Max-Planck-Gesellschaft and the Heidenhain-Stif\-tung.

\end{acknowledgements}

\end{document}